\documentclass{PMIDIC2024}

\title{Training an AI hyperelastic constitutive model with experimental data }
\author{\underline{C. Jailin}, A. Benady and E. Baranger}
\address{Universit\'e Paris-Saclay, CentraleSup\'elec, ENS Paris-Saclay, CNRS, \\ LMPS -- Laboratoire de M\'ecanique Paris-Saclay, 91190, Gif-sur-Yvette, France.}
\begin{document}
% ---------------------------------------------------------------------
\maketitle
% ---------------------------------------------------------------------

\begin{abstract}
A Physics-Augmented Neural network is trained to model a hyperelastic behavior. The dataset used for the training, validation, and test are displacement-force couples obtained from two experiments on a rubber-like material. One experiment was dedicated for the test, to assess the capacity of the model to generalize on unseen loadings and geometries. The trained AI model outperforms a standard Neo Hookean model identified on the same data. Particular attention is paid to the mechanical data information contained in the different datasets.
\end{abstract}

\paragraph{Introduction}
Recent advancements in artificial intelligence and experimental mechanics have paved the way for innovative approaches to model material behaviors. By enforcing physics and mechanical information in the neural network, their efficiency has greatly improved, allowing them to learn complex behaviors while requiring a reduced number of data. Those mechanically AI-informed constitutive models are called Physics-Augmented Neural Networks (PANN)~\cite{linden2023neural}.
Showing promising results from synthetic (yet realistic) experiments, obtained from an initial FEA simulation, those physics-augmented constitutive models have not been trained on real experimental data. Being real data, the true material behavior is unknown (which complexifies the model design), the data are affected by experimental noise/uncertainties, and the final evaluation cannot be challenged to ground truth. It is here presented the optimization and training of a hyperelastic PANN model from DIC and force measurement data.

\paragraph{Methods}

\begin{figure}[h]
\centering
     {\includegraphics[width=1.\textwidth]{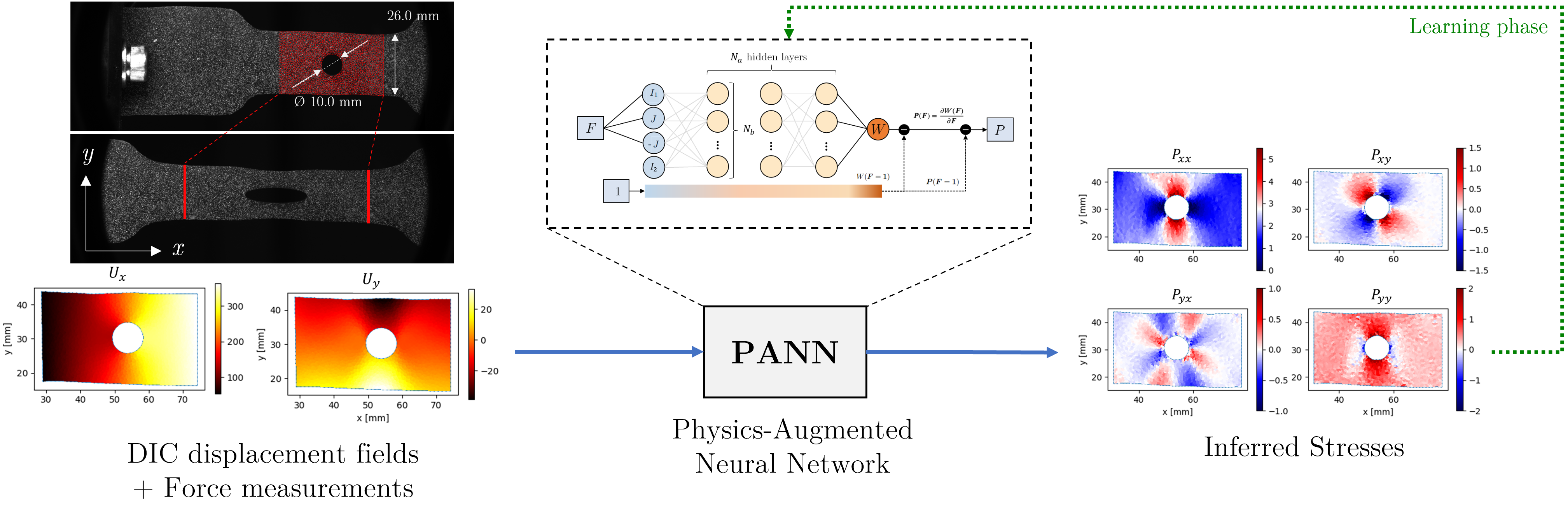}}
	\caption{Constitutive model based on a Physics-Augmented Neural Network}
	\label{fig:PANN} 
\end{figure}

\emph{PANN constitutive model}: The constitutive model used in this study~\cite{jailin2024} is based on a Physics-Augmented Neural Network. 
Multiple physical constraints are enforced on the neural network (e.g., convexity and positivity of the free energy) to ensure a thermodynamically relevant strain ($\bm F$) to stress ($\bm P$) mapping (Figure~\ref{fig:PANN}).  
The model $\mathcal{M}_{\text{PANN}}$ is trained in an unsupervised framework (so-called NN-EUCLID~\cite{thakolkaran2022nn}), minimizing a loss function $\mathcal{L}[\mathcal{M}(\bm F),\bm R]$ ensuring the whole sample is balanced considering the measured reaction forces $\bm R$. The network weights $p$ are finally optimized using gradient-based techniques: $\hat{p}=\underset{p}{\text{Argmin}}  \mathcal{L} [\mathcal{M}_{\text{PANN}}(p,\bm F),\bm R]$. 

Because the true behavior is unknown, 10 different model architectures, comprising 145 up to 17,217 trainable parameters, were evaluated to select the most accurate. Finally, the PANN model is compared with a standard Neo-Hookean law.

\emph{Dataset}: The study uses two distinct uni-axial experiments on a rubber-like material, achieving local deformations over 200\% and exhibiting a highly non-linear behavior. Full-field measurements were extracted using global DIC approaches with brightness-contrast corrections~\cite{sciuti2021benefits}. Together with axial force measurements, this constitutes a dataset of 661 deformation fields/force couples $\{\bm F,\bm R\}$ for each loading step and geometry. Data from these experiments were carefully split into 20 training steps (to learn the model weights), 6 validation steps (to monitor the model performances, and optimize the model architecture), and 635 test steps (for the final evaluation and metrics computation).

\paragraph{Results}
The different model architectures are compared on the validation dataset and evaluated through the equilibrium gap loss, allowing for the identification of the best model hyperparameters (\textit{e.g.} number of layers, number of neurons). The chosen model is then compared to a Neo Hookean model identified using a similar process. Based on the results of different metrics (inner and boundary equilibrium), the PANN model performs in the same range if not better than the Neo Hookean model. On the experiment dedicated to the test, the PANN model outperforms the Neo-Hookean law.
The PANN model does not show discontinuities at the trained loading steps and provides accurate results beyond the trained steps. Although its mechanical content may be low informative in the unseen experiment, the PANN model performs better than the Neo Hookean model, showing its generalization capacity. 

\paragraph{Discussion and Conclusion} 
The PANN model showed its capacity to learn the material's hyperelastic behavior, outperforming traditional Neo-Hookean models. It demonstrated robustness not only within the range of trained loads but also in unseen loadings, illustrating a capacity for both interpolation and extrapolation (at least on the evaluated states).

%EB : finalement tu ne parles pas beaucoup du contenu informationnel. vu la place restante (si c'est 2pg maxi), une phrase d'ouverture devrait suffire
The results underscore the importance of carefully controlling the datasets used for the training, validation, and test. The mechanical content in those subsets has to be informative enough for the training to learn the constitutive model from a large range of mechanical states, and challenging enough for the test to provide an accurate evaluation. Although performed with two geometries and large deformations, the two experiments used in the analysis do not exhibit complex stress/strain multi-axial states, thus limiting the potential for learning. 

Collecting vast amounts of data from simple mechanical tests alone is not enough; instead, emphasizing the informative mechanical content of that data is crucial for the effective training of the network.
Quantifying, comparing, visualizing, and even optimizing the data contained in mechanical tests hence remains a crucial challenge.

% ---------------------------------------------------------------------
%
% ---------------------------------------------------------------------
% -----------------------------------

% ---------------------------------------------------------------------
\end{document}